\documentclass[aps, prl, reprint, superscriptaddress]{revtex4-1}

\usepackage{amsmath}
\usepackage{amssymb}
\usepackage{wasysym}
\usepackage{graphicx}
\usepackage{hyperref}
\usepackage{color}
\usepackage{physics}
\usepackage{siunitx}
\usepackage{xcolor}
\usepackage{changes}
\usepackage{comment}
\usepackage{siunitx}
\usepackage{bm}
\usepackage[normalem]{ulem}

\allowdisplaybreaks

\begin{document}

\title{Transition from a polaronic condensate to a degenerate Fermi gas \\ of heteronuclear molecules}

\author{Marcel Duda}
\author{Xing-Yan Chen}
\author{Andreas Schindewolf}
\author{Roman Bause}
\author{Jonas von Milczewski}
\author{Richard Schmidt}
\affiliation{Max-Planck-Institut f\"{u}r Quantenoptik, 85748 Garching, Germany}
\affiliation{Munich Center for Quantum Science and Technology, 80799 M\"{u}nchen, Germany}
\author{Immanuel~Bloch}
\affiliation{Max-Planck-Institut f\"{u}r Quantenoptik, 85748 Garching, Germany}
\affiliation{Munich Center for Quantum Science and Technology, 80799 M\"{u}nchen, Germany}
\affiliation{Fakult\"{a}t f\"{u}r Physik, Ludwig-Maximilians-Universit\"{a}t, 80799 M\"{u}nchen, Germany}
\author{Xin-Yu~Luo} \email{Corresponding author: \\E-Mail: xinyu.luo@mpq.mpg.de}
\affiliation{Max-Planck-Institut f\"{u}r Quantenoptik, 85748 Garching, Germany}
\affiliation{Munich Center for Quantum Science and Technology, 80799 M\"{u}nchen, Germany}

\date{\today}

\begin{abstract}
The interplay of quantum statistics and interactions in atomic Bose--Fermi mixtures leads to a phase diagram markedly different from pure fermionic or bosonic systems. However, investigating this phase diagram remains challenging when bosons condense. Here, we observe evidence for a quantum phase transition from a polaronic to a molecular phase in a density-matched degenerate Bose--Fermi mixture. The condensate fraction, representing the order parameter of the transition, is depleted by interactions and the build-up of strong correlations results in the emergence of a molecular Fermi gas. By driving through the transition, we ultimately produce a quantum-degenerate sample of sodium-potassium molecules exhibiting a large molecule-frame dipole moment of 2.7 Debye. The observed phase transition represents a new phenomenon complementary to the paradigmatic BEC-BCS crossover observed in Fermi systems.

\end{abstract}

\maketitle

\section{Introduction}

Mixtures of interacting bosons and fermions have been subject of intense research since conventional superconductivity was understood to arise from the effective attraction between electrons mediated by phonons. In solid-state materials, the electron-phonon coupling is captured by Fr\"ohlich or Holstein models. Developments in ultracold atoms \cite{chin2010feshbach} and van-der-Waals materials \cite{Sidler2017} now make it possible to realize Bose--Fermi mixtures that are governed by beyond-Fr\"ohlich physics where bosons and fermions can bind to fermionic molecules \cite{Ni2008,Park2015,Rvachov2017,Seesselberg2018,Yang2019,mak2013tightly} reaching the quantum-degenerate regime \cite{DeMarco2019}.  The competition between this novel bound state physics and mediated interactions leads to an enriched phase diagram potentially featuring supersolidity and charge-density-wave phases  \cite{Enss2009,Matuszewski2012,Shelykh2010,Cotlet2016}, molecular Fermi liquids \cite{Powell2005, Suzuki2008,Fratini2010,Ludwig2011,Guidini2015,vonMilczewski2021}, and unconventional boson-induced superconductivity \cite{Kinnunen2018,Laussy2010}.

Experimentally, the phase diagram of Bose--Fermi mixtures has primarily been explored in the regime of large population imbalance where one species acts as a dilute, thermal gas of impurities dressed by its environment.  Bose polarons were recently observed in the limit of fermionic impurities in a bosonic bath \cite{Hu2016,yan2020bose}, while the existence of a transition from Fermi polarons to molecules has by now been firmly established for impurities immersed in a Fermi sea \cite{Schirotzek2009,Koschorreck2012,Ness2020,Fritsche2021}. However, so far it has remained unclear how the transition from atoms to molecules proceeds when the impurities are degenerate, in particular when bosons and fermions of comparable density dress each other mutually and drastically modify their respective behavior. Importantly, this regime of matched particle density is promising for the association of heteronuclear molecules at high phase space density, which finds wide-ranging applications in quantum chemistry and the exploration of dipolar quantum many-body systems \cite{Lincoln2009,bohn2017}. At the same time, reaching this regime is notoriously difficult in double-degenerate mixtures due to the enhanced density of the bosonic condensate. The excess density of the Bose--Einstein condensate (BEC) causes fast interspecies loss which remains the key bottleneck for reaching quantum degeneracy in heteronuclear molecules and for the study of strongly correlated Bose--Fermi mixtures.

\begin{figure*}[t]
\centering
\includegraphics{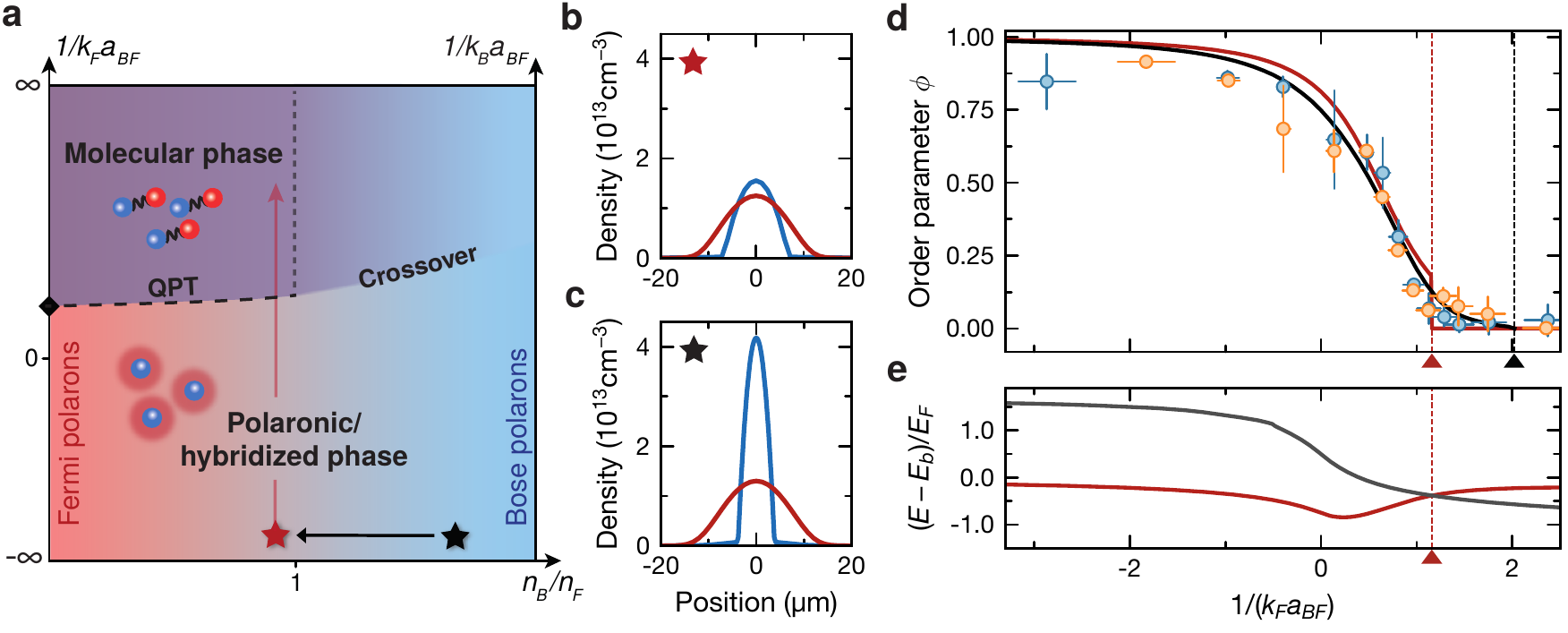}
\caption{\textbf{Quantum phase transition in a density-matched Bose--Fermi mixture.}
(a) Phase diagram of degenerate Bose--Fermi mixtures as a function of the density ratio $n_B/n_F$ and the dimensionless interaction strength $(k_i a_{BF})^{-1}$. For $n_B/n_F\to 0$ one attains the Fermi-polaron limit featuring a polaron-to-molecule transition (black diamond), while for $n_B/n_F\to \infty$ the Bose-polaron limit with a smooth crossover is reached. (b) Calculated in-situ density profiles of bosons (blue) and fermions (red) in the species-dependent 785-nm dipole trap, and  (c) in the far-detuned trap \cite{supp}. (d) Order parameter $\phi$ as a function of $(k_F a_{BF})^{-1}$ for the boson-fermion peak-density ratio $\tilde{n}_B/\tilde{n}_F=0.9$ (orange points) and $\tilde{n}_B/\tilde{n}_F=1.3$ (blue points). The error bars are discussed in \cite{supp}. The black solid line shows the order parameter  from zero-temperature theory in Ref.~\cite{Guidini2015} predicting the QPT to occur at $(k_F a_{BF})_c^{-1}=2.02$ (black triangle) for ideal bosons interacting with a Fermi gas at $n_B/n_F=1$. The red solid line shows the polaron quasiparticle weight of a bosonic impurity in a Fermi gas obtained from a self-consistent functional renormalization group (fRG) calculation that predicts the polaron-to-molecule transition to occur at $(k_F a_{BF})_c^{-1}=1.16$ (red triangle). (e) Energy spectrum of the zero-momentum Fermi polaron (red line) and the zero-momentum molecule (gray line) for a single bosonic impurity obtained from the fRG. The energies cross at the polaron-to-molecule transition (red dashed line). For $a_{BF}>0$, the binding energy $E_b=\hbar^2 / 2 \mu a^2_{BF}$ is subtracted where $\mu$ is the reduced mass.
}
\label{fig:qpt}
\end{figure*}

In this work, we show that in the low-temperature regime where the bosonic impurities condense, strong boson-fermion interactions drive a quantum phase transition (QPT) from a polaronic condensate to a molecular Fermi gas. Using a novel density-decompression technique which mitigates atomic loss, we produce a double-degenerate Bose--Fermi mixture of $\mathrm{^{23}Na}$ and $\mathrm{^{40}K}$ with matched density and reveal signatures of the QPT. Starting from a weakly interacting mixture, increased attractive interactions dress the bosonic condensate polaronically. By continuously tuning the interaction strength, the polaronic condensate is depleted and a transition into a phase of quantum-degenerate fermionic molecules is observed. Driving the underlying QPT enhances the association efficiency of Feshbach molecules to near unity. This enables the subsequent creation of ground-state $\mathrm{^{23}Na^{40}K}$ molecules with a large molecular-frame dipole moment of 2.7 Debye in the quantum-degenerate regime.

\section{Degenerate Bose--Fermi mixtures in the density-matched regime}
A simplified phase diagram of Bose--Fermi mixtures is illustrated in Fig.~\ref{fig:qpt}a as a function of the ratio of boson to fermion density $n_B/n_F$ and the dimensionless interaction strength $1/k_i a_{BF}$. Here, $a_{BF}$ denotes the boson-fermion scattering length and the wave vector $k_i$ is determined by the interparticle spacing of the majority species $k_i = (6 \pi^2 n_i)^{1/3}$ where i denotes B(F) for $n_B>n_F$ ($n_B<n_F$). We only consider phases at sufficiently high temperatures where instabilities towards phases such as charge-density waves and superfluid fermion pairing can be ignored. Phases involving bound states of more than one boson are also ignored as these are intrinsically unstable due to fast recombination loss.

Qualitatively, the phase diagram can then be divided into two regimes. In the limit of vanishing interactions, $(k_i a_{BF})^{-1} \to -\infty$, bosons and fermions decouple. As attractive interactions are switched on, fermions and bosons modify each other's properties, leading to quasiparticle formation. Due to the polaronic character of this interaction, we denote the resulting phase as the \emph{Polaronic phase}. In the limit of strong attraction, i.e. $(k_i a_{BF})^{-1} \to \infty$, and for $n_B\leq n_F,$ binding of all bosons to fermions leads to a Fermi sea of molecules coexisting with an atomic Fermi sea; we denote this phase as the \emph{Molecular phase}. For $n_B\leq n_F$ these two phases are predicted to be separated by a quantum phase transition \cite{Fratini2010,Guidini2015,Powell2005,Ludwig2011} (black dashed line in Fig.~\ref{fig:qpt}a), while for $n_B > n_F$ a crossover is expected due to the presence of an excess condensate \cite{Powell2005,Ludwig2011}. 
When tuning the density ratio across $n_B/n_F\approx 1$  in the regime of strong attraction, an additional crossing of the phase transition (gray dotted line in Fig.~\ref{fig:qpt}a) is predicted to occur \cite{Powell2005} whose nature might be of first-order \cite{Ludwig2011}. Most experiments have been carried out either on the far left- or the far right-hand side of the phase diagram. 

A natural way to investigate the phase diagram away from these impurity limits starts from producing a double-degenerate Bose--Fermi mixture. Especially the regime of matched densities is of interest where the system becomes strongly correlated and neither of the atomic species can be regarded as a quantum impurity. To access this novel regime, we employ a species-dependent dipole trap at 785 nm, which is near-detuned to the $D$-lines of the K atoms. This trap provides a weaker confinement of the Na compared to the K atoms, lowering the density of the Na BEC and increasing overlap between the species (see Fig.~\ref{fig:qpt}b). As a consequence, the detrimental loss resulting from collisions of Na atoms in the BEC with $\mathrm{NaK^{*}}$ Feshbach molecules is dramatically reduced. In contrast, for a typical trap setup where the trapping effect is similar for both atomic species, the peak density of the BEC is considerably larger than that of the Fermi gas (see Fig.~\ref{fig:qpt}c). This results in an entirely different physical regime related to the Bose-polaron problem with a low molecule association efficiency when starting from the BEC \cite{supp}.

\section{QPT from a polaronic phase to a molecular phase}
In the regime where $n_B\leq n_F$, theory predicts a transition from a phase where a BEC coexists with a Fermi gas to a liquid where all bosons are bound into molecules \cite{Guidini2015,Powell2005}. By tuning the interactions from weak to strong coupling, the boson-fermion interaction gradually depletes the BEC until the condensate fraction, representing the order parameter of the transition, vanishes at the critical interaction parameter $(k_F a_{BF})_c^{-1}$. At this point a quantum phase transition, possibly masked by a narrow phase-separation regime \cite{Ludwig2011,Bertaina2013}, occurs. This is a distinct feature in Bose--Fermi mixtures that is absent for the BEC-BCS crossover in spin-balanced fermionic mixtures where no symmetry-breaking pattern is changed as the interaction strength is varied across the Feshbach resonance~\cite{zwerger2011bcs}. 

It is predicted that the depletion of the condensate depends weakly on the boson-fermion density ratio which, remarkably, extends all the way to the Fermi polaron limit, $n_B/n_F \to 0$ \cite{Guidini2015}. In the extreme limit of a single bosonic impurity in a Fermi gas, the phase transition connects to a polaron-to-molecule transition and the condensate fraction reduces to the impurity quasiparticle weight \cite{Guidini2015}. A computational, self-consistent  functional renormalization group (fRG) technique that takes into account an infinite number of particle-hole excitations of the Fermi sea \cite{Schmidt2011,supp} predicts this transition to occur at  $(k_F a_{BF})^{-1}=1.16$ (see Fig.~\ref{fig:qpt}e). This transition point is expected to shift to larger values as temperature and boson-density increase \cite{Guidini2015,Ness2020,Parish2021}.

\begin{figure*}
\centering
\includegraphics{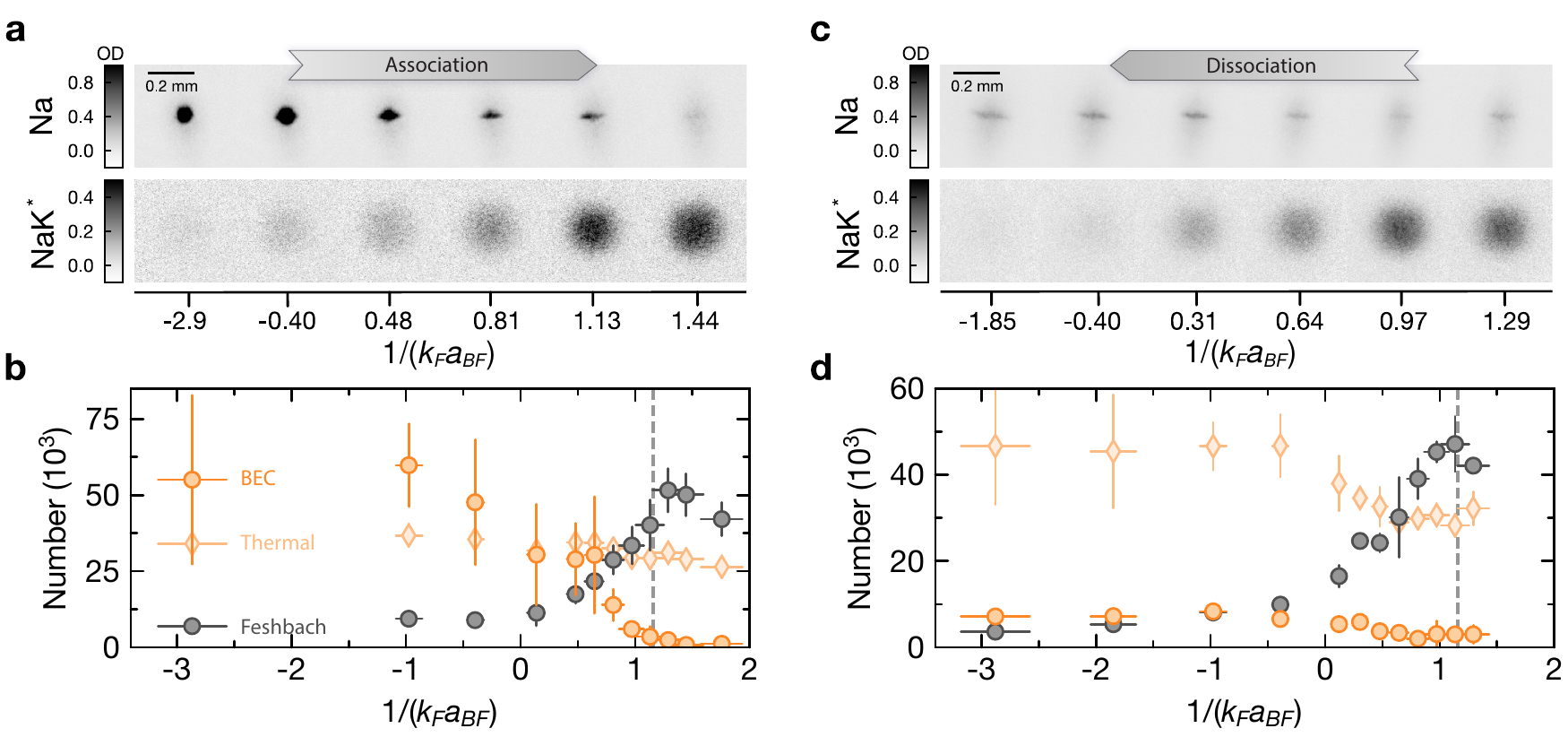}
\caption{\textbf{Association and dissociation process of degenerate Feshbach molecules.} (a) Absorption images of Na atoms (Na) and Feshbach molecules ($\mathrm{NaK^*}$) after \SI{18}{ms} time of flight during the association ramp from the polaronic BEC to the molecular phase. (b) Production of Feshbach molecules. Number of condensed Na atoms (dark orange points), thermal Na atoms (bright orange diamonds) and Feshbach molecules (gray points) is shown as a function of $(k_F a_{BF})^{-1}$ for $\tilde{n}_B/\tilde{n}_F=1.3$. The gray line indicates the polaron-to-molecule transition at $(k_F a_{BF})^{-1}=1.16$ in the Fermi polaron problem. (c) Absorption images during the dissociation ramp with \SI{18}{ms} time of flight. (d) Dissociation of Feshbach molecules. Condensed Na atoms (dark orange points), thermal Na atoms (bright orange diamonds) and Feshbach molecules (gray points) are shown as a function of $(k_F a_{BF})^{-1}$ for $\tilde{n}_B/\tilde{n}_F=1.3$. The error bars are discussed in \cite{supp}.}
\label{fig:association_dissociation}
\end{figure*}

In the following, we probe signatures of this QPT. Our experiment typically starts with $2.3 \times 10^5$ $\mathrm{^{40}K}$ atoms at a temperature $T=80$\,nK (corresponding to $T/T_F\sim 0.2$) and a BEC of $0.8 \times 10^5$ $\mathrm{^{23}Na}$ atoms with a condensate fraction of $60\%$ at a magnetic field of \SI{81}{G}. We employ a single magnetic field ramp with a speed of \SI{3.5}{G/ms} \cite{supp} that is terminated at the desired magnetic field close to a Feshbach resonance at \SI{78.3}{G} \cite{Chen2021}, corresponding to different interaction strengths $(k_F a_{BF})^{-1}$.
We checked that the depletion of the BEC is independent of the ramp speed for the sufficiently slow ramp used here \cite{supp}. We then quench the magnetic field to \SI{72.3}{G}. This projects the system onto free atoms and deeply bound molecules which are subsequently imaged in time of flight after Stern--Gerlach separation as shown in Fig.~\ref{fig:association_dissociation}a, b.

To characterize the phase transition quantitatively, we define the normalized order parameter $\phi = N_{\mathrm{BEC}}/(N_{\mathrm{m}}+N_{\mathrm{BEC}})$. Here $N_{\mathrm{BEC}}$ and $N_{\mathrm{m}}$ represent the number of condensed Na atoms and those associated to molecules, respectively. The order parameter $\phi$ describes the depletion of the condensate fraction due to the excitation of bosons to finite-momentum states by quantum fluctuations. These quantum fluctuations are dominated by the build-up of pairing correlations measured by the projection onto molecules. 
Note, in the definition of $\phi$ we disregard thermal Na atoms whose number remains mostly unchanged across the full interaction range. Thus, we may directly compare our findings with predictions from zero-temperature theory as shown in Fig.~\ref{fig:qpt}d. Specifically, we make a comparison with the quasiparticle weight of a bosonic impurity calculated from a self-consistent fRG approach \cite{Schmidt2011,supp} (red dashed line) and predictions of the condensate fraction from a non-self-consistent T-matrix (NSCT) theory \cite{Guidini2015} (black solid line) for $n_B=n_F$ which neglects multiple particle-hole excitations in the Fermi sea.

In Fig.~\ref{fig:qpt}d, we show the measured order parameter $\phi$ as a function of $(k_F a_{BF})^{-1}$ for $\tilde{n}_B/\tilde{n}_F = 0.9$ and $1.3$ where $\tilde{n}_B/\tilde{n}_F$ denotes the boson-fermion peak-density ratio. As the interaction strength increases, $\phi$ reduces slowly for $(k_F a_{BF})^{-1} < 0$. However, once the scattering length becomes positive, $\phi$ decreases rapidly and vanishes in the regime beyond $(k_F a_{BF})^{-1}=1.43(15)$ where the residual condensate fraction is comparable to the uncertainty of the measurement. Our measurements agree well with the predicted condensate fraction from the NSCT approach throughout the entire interaction regime. Importantly,  both data sets overlap within error bars,
providing support for the predicted universality of the condensate depletion with respect to varying $n_B/n_F$ \cite{Guidini2015}. This, in turn, justifies comparing our in-trap experiment with predictions for a homogeneous system. The data also show a remarkable agreement with the quasiparticle weight of a single impurity in most of the interaction regime except close to the phase transition. Here the order parameter vanishes smoothly in contrast to a jump expected in the impurity limit. This indicates that, despite having a large boson density, the system can be well described as a condensate of polaronically dressed bosons.

To estimate the transition point independently from the slowly varying order parameter, we consider the projected Feshbach molecule number shown in Fig.~\ref{fig:association_dissociation}b as a measure of existing boson-fermion pairing correlations \cite{supp}. As $(k_F a_{BF})^{-1}$ increases, so do the  pairing correlations (including potential fermion-molecule mixing \cite{Powell2005}) until they saturate when the bosons are fully bound into molecules. The resulting transition point $(k_F a_{BF})^{-1}=1.29(13)$ extracted from the measured Feshbach molecule number is consistent with the transition point obtained through the vanishing of the order parameter.

Driving the QPT provides an efficient method to create molecules. Our data show that a striking conversion efficiency of around $80\%$ of the Na atoms in the BEC into Feshbach molecules can be achieved. We believe that for perfect density matching between the BEC and the Fermi gas, the entirety of the BEC can be converted into molecules. In contrast, the highest conversion efficiency from a BEC previously reported was less than $50\%$ for a density ratio of $\tilde{n}_B/\tilde{n}_F \approx 10$ in $\mathrm{KRb}$ \cite{DeMarco2019}, which was  possible because of the ten-fold lower interspecies loss coefficients compared to $\mathrm{NaK}$~\cite{DeMarco2019,Chen2021,Bloom2013}. 

Next, we investigate the reversal of the phase transition. After the association ramp reaches $(k_F a_{BF})^{-1}=1.3$, the magnetic field is ramped back to dissociate the molecules. The dissociation ramp is again followed by a magnetic-field quench to \SI{72.3}{G} for detection. As can be seen from the time-of-flight images in Fig.~\ref{fig:association_dissociation}c, the number of projected Feshbach molecules decreases, while a finite BEC fraction is recovered. In particular, we show that the number of Na atoms in the BEC can be increased from $3(2)\times10^3$ to $8(1) \times10^3$ (see Fig.~\ref{fig:association_dissociation}d). Heating is evident after the dissociation in the form of an increase of thermal Na atoms which we attribute to the non-adiabatic nature of the magnetic field ramps near the transition point. Due to the changing number of Na atoms in the thermal wings, we thus cannot characterize the reversal of the phase transition with the order parameter $\phi$ as done in the association ramp. Nonetheless, the partial restoration of the BEC highlights the coherence preserved in our experiment and is a striking example of how bosons that were bound to fermionic molecules in finite-momentum states are converted back into their motional ground state. In addition, the restoration necessitates the existence of a deeply quantum-degenerate gas of Feshbach molecules and underlines the low entropy of our molecular clouds.

\begin{figure}
\centering
\includegraphics{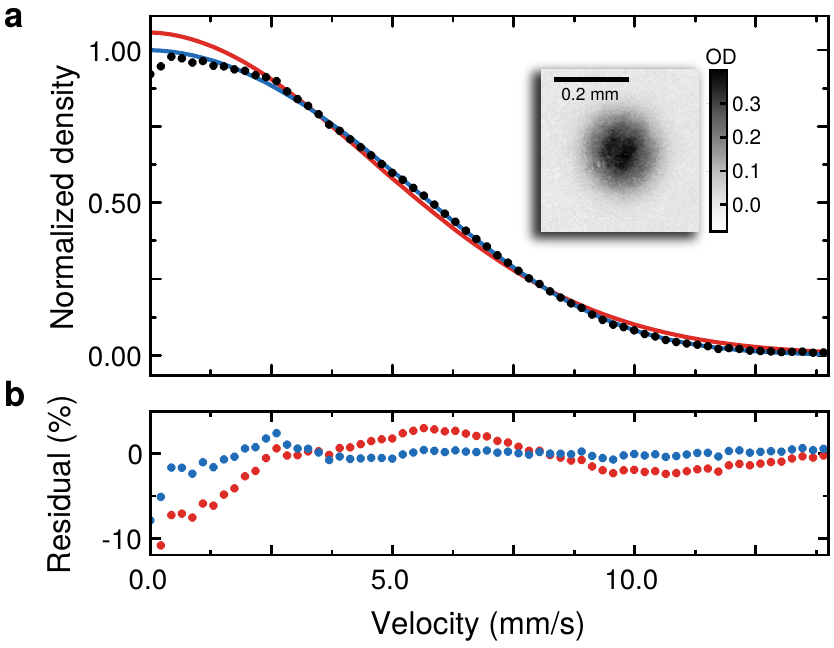}
\caption{\textbf{Quantum degeneracy of Feshbach molecules.} (a) Velocity distribution of Feshbach molecules. The azimuthal integral (black points) of an average of fifteen images with a time of flight of \SI{15}{ms} (shown in the inset) was fitted with a Fermi--Dirac (blue line) and a Gaussian distribution (red line). Compared to the Gaussian fit, the data shows a higher occupation at larger momenta which is characteristic for the Fermi pressure in a degenerate Fermi gas. The fit of the Fermi--Dirac distribution results in a $T/T_F = 0.28(1)$. (b) Azimuthal integral of the residuals.}
\label{fig:degeneracy}
\end{figure}

\section{Degenerate Fermi gas of $\mathbf{NaK^*}$ Feshbach molecules}
After the preparation $5\times10^4$ of Feshbach molecules at a temperature of 100\,nK by ramping across the QPT, we convert them to more deeply bound states by quenching the magnetic field to \SI{75}{G}. At this magnetic field, we turn on a strong gradient to levitate the molecules and remove any residual atoms from the trap \cite{supp}. Inelastic collisions between molecules are strongly suppressed by Pauli-blocking leading to a second-long lifetime \cite{supp}. The quantum degeneracy of the molecular gas after association is determined by time-of-flight imaging after holding the molecules for \SI{100}{ms} (see Fig.~\ref{fig:degeneracy}), ensuring that collective oscillations induced by the magnetic field ramps of the molecular cloud are dampened out \cite{supp}. The momentum distribution of the molecules is well described by a Fermi--Dirac distribution \cite{DeMarco2019} with a temperature of $T=0.28(1)~T_F$.

\section{Low-entropy $\mathbf{NaK}$ ground-state molecules}

Having established a means to efficiently create large ensembles of heteronuclear Feshbach molecules sets the stage for the creation of polar ground-state molecules. To achieve this, we coherently transfer the degenerate Fermi gas of $5\times10^4$ Feshbach molecules into the rovibronic ground state of the $X^1\Sigma^+$ electronic manifold \cite{Bause2021b} with an overall efficiency of up to $60\%$. The momentum distribution is maintained during the transfer and is still fit well by a Fermi--Dirac distribution at a temperature of $T=0.28(3)~T_F$. However, due to the absence of thermalization and the imperfect transfer efficiency, random holes are created in the molecular Fermi sea. Thus we expect the quantum degeneracy of the ground-state molecular gas to be reduced compared to the Feshbach molecular gas \cite{DeMarco2019,Tobias2020}. Hence, as an alternative measure of the degeneracy we compute the peak-occupancy of the Fermi gas $f$, characterizing the probability of the lowest energy state in the Fermi gas to be occupied. Based on the transfer efficiency to the ground state and the initial degeneracy of the Feshbach molecules, we obtain a peak occupancy of $f=57(3)\%$ which corresponds to an \emph{effective} temperature of $T=0.52(2)~T_F$.

\section{Conclusion}
Using a species-dependent decompression technique of atomic clouds, we have established the existence of a QPT in Bose--Fermi mixtures between a phase featuring condensation and a molecular Fermi gas in excellent quantitative agreement with theory. By driving the system through this QPT, we have produced a gas of quantum-degenerate Feshbach molecules with a record efficiency. The observed phase transition represents a new phenomenon complementary  to the paradigmatic BEC-BCS crossover observed in Fermi systems \cite{zwerger2011bcs} and the atomic-to-molecular BEC crossover in Bose systems \cite{zhang2021transition}. It is a first step in the exploration of strong-correlation physics in degenerate Bose--Fermi systems and provides a benchmark for their theoretical understanding. Our method can be readily extended to other Bose--Fermi mixtures for producing large degenerate samples of fermionic molecules and may help to achieve a heteronuclear molecular BEC from Bose--Bose mixtures which suffer from even more severe losses when both bosonic species condense \cite{Takekoshi2014,Molony2014,Guo2016,Voges2020,Cairncross2021,Warner2021}. Ultimately our technique allows us to produce a gas of nonreactive ground-state molecules in the quantum-degenerate regime with five times stronger molecular-frame dipole moments than the first degenerate polar molecules of KRb. This opens up exciting opportunities to study strongly correlated dipolar quantum systems \cite{baranov_condensed_2012} ranging from the collapse of dipolar Fermi gases \cite{Vladimir2019} to extended Heisenberg XXZ models \cite{Buchler_spinmodel_2012} and extended Fermi--Hubbard models \cite{Bruun_stripephase_2012} in optical lattices.

\section{Acknowledgments}
%\begin{acknowledgments}
We thank P. Pieri for stimulating discussions and providing the calculations of the condensate fraction for a equal-density mixture, T. Shi for stimulating discussions, X. Li and T. Hilker for insightful comments on our manuscript. We thank the previous members of the MPQ NaK Lab, especially C. Gohle, for their contributions to the experimental apparatus. \textbf{Funding:} We acknowledge support by the Deutsche Forschungsgemeinschaft under Germany's Excellence Strategy – EXC-2111 – 390814868 and under Grant No. FOR 2247. The experimental team acknowledges support from the Max Planck Society and the European Union (PASQuanS Grant No. 817482). J.v.M. is supported by a fellowship of the International Max Planck Research School for Quantum Science and Technology (IMPRS-QST). A.S. acknowledges funding from the Max Planck Harvard Research Center for Quantum Optics. \textbf{Author contributions:} All authors contributed substantially to the work presented in this manuscript. M.D. and X-Y.C. carried out the experiments and, together with A.S. and R.B., they maintained and improved the experimental setup.  M.D. analyzed the data. J.v.M. and R.S. performed the theoretical calculations. R.S., I.B. and X-Y.L. supervised the study. All authors worked on the interpretation of the data and contributed to the final manuscript. \textbf{Competing interests:} The authors declare no competing interests.
%\end{acknowledgments}

\bibliography{bibliography}

\clearpage

\section*{Supplemental Material for "Transition from a polaronic condensate to a degenerate Fermi gas of heteronuclear molecule"}

\subsection{Preparing a density-matched degenerate Bose--Fermi mixture}
\begin{figure}
\centering
\includegraphics{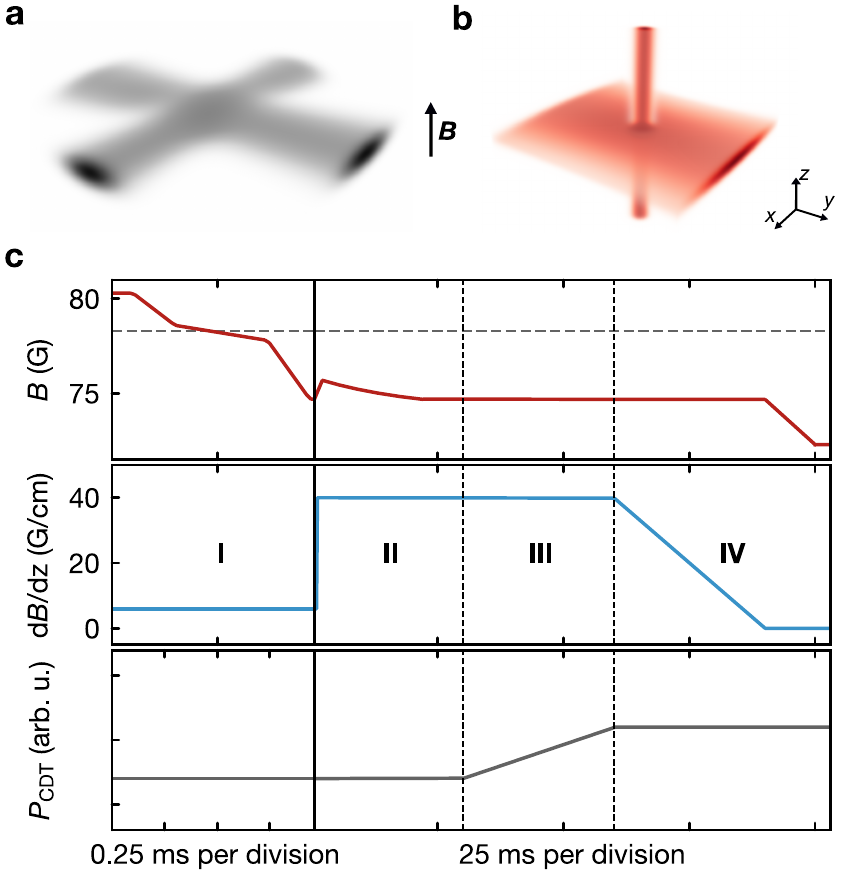}
\caption{Trap configurations and sequence for molecule association. (a) Trap configuration of the 1550/1064-nm trap and (b) the 785-nm trap. The magnetic field and its gradient are along the vertical direction (z-direction). (c) Typical timing sequence for the association of ground-state molecules showing ramps of the magnetic field (red), its gradient (blue) and the power of the 1550/1064-nm trap (gray). I. Association of Feshbach molecules by three-staged magnetic field ramp. II. Removal of unassociated atoms. III. Compression of the 1550/1064-nm trap while the power of the 785-nm trap remains unchanged. IV. Removal of the levitation gradient for subsequent STIRAP pulses. The black vertical solid and dashed lines separate different stages of the sequence. The horizontal gray dashed line marks the Feshbach resonance at \SI{78.3}{G}.}
\label{fig:sequence}
\end{figure}

$\mathrm{^{23}Na}$ atoms and $\mathrm{^{40}K}$ atoms are loaded into a 3D magneto-optical trap (MOT) from a Zeeman-slower stage for $\mathrm{^{23}Na}$ and a 2D-MOT stage for $\mathrm{^{40}K}$, followed by gray-molasses cooling for Na atoms on the $D_\mathrm{1}$ line. Both species are transferred into an optically plugged magnetic quadropole trap and are cooled by forced radio-frequency evaporation. After evaporation, there are $1.6 \times 10^8$ Na atoms and $1.0 \times 10^6$ K atoms at a temperature of \SI{6}{\mu K}, of which $2 \times 10^7$ Na atoms and $6 \times 10^5$ K atoms are loaded into an optical trap that transports the mixture from the MOT chamber to a glass cell. This transport trap is formed by two 1064-nm laser beams that intersect at an angle of $3 ^{\circ}$ to increase axial confinement \cite{Gross2016}. After the transport, the mixture is transferred into a crossed optical dipole trap formed by one of the transport beams and an additional 1550-nm beam (1550/1064-nm trap, see Fig.~\ref{fig:sequence}a). The waists of these two beams in vertical and horizontal direction are $50 \times 100~\mathrm{\mu m}$ and $50 \times 150~\mathrm{\mu m}$, respectively. Optical evaporation is performed by exponentially ramping down the beam powers. During the evaporation, a trap formed by two 785-nm beams with waists of $30 \times 400~\mathrm{\mu m}$ and $50 \times 50~\mathrm{\mu m}$ (785-nm trap, see Fig.~\ref{fig:sequence}b) is ramped on. The wavelength of this trap is chosen such that it is closely detuned from the $D$-lines of the K atoms and the polarizabilities of the Feshbach and ground-state molecules are equal \cite{vexiau2017dynamic}. After evaporation, the K atoms undergo two Landau-Zener sweeps for a hyperfine transfer from $|F,m_F \rangle=|9/2,9/2 \rangle$ to $|9/2,-9/2 \rangle$, where $F$ is the  total  atomic  angular  momentum  and $m_F$ is its projection onto the axis of the magnetic field. The Na atoms remain in $|1,1 \rangle$. The first state transfer from $|9/2,9/2 \rangle$ to $|9/2,-7/2 \rangle$ is performed at a magnetic field of \SI{74}{G} and the second state transfer from  $|9/2,-7/2 \rangle$ to $|9/2,-9/2 \rangle$ is performed at \SI{81}{G}. The magnetic fields are chosen to minimize changes in the interspecies interaction during the sweeps. After the state preparation, we turn on a magnetic field gradient of \SI{6.4}{G/cm} to levitate the Na atoms and K atoms against gravity. Afterwards the 1550/1064-nm trap is ramped down, decompressing mainly the Na atoms and leading to a trap configuration that is dominated by the 785-nm beams. In this trap configuration, the trap frequencies of the K atoms are approximately three times larger than the ones of the Na atoms. This procedure leads to the preparation of a density-matched mixture of $2.3 \times 10^5$ $\mathrm{^{40}K}$ atoms at a temperature $T=80$\,nK (corresponding to $T/T_F\sim 0.2$, where $T_F$ is the Fermi temperature) and a BEC of $0.8 \times 10^5$ $\mathrm{^{23}Na}$ atoms with a condensate fraction of about $60\%$.

\subsection{Feshbach association and preparation for ground-state transfer}
To associate Feshbach molecules, we employ a series of magnetic-field ramps (see Fig.~\ref{fig:sequence}b). First, the magnetic field is quickly ramped to \SI{78.6}{G}, followed by the association ramp across the Feshbach resonance at \SI{78.3}{G} with a ramp speed of \SI{3.5}{G/ms}. This slower ramp stops at \SI{77.8}{G} where the formation of Feshbach molecules saturates. In order to minimize the loss of the Feshbach molecules, we subsequently ramp the magnetic field quickly to \SI{75}{G}. At this magnetic field, the magnetic moment of the Feshbach molecules reduces by a factor of six compared to the open-channel dominated Feshbach molecules close to the resonance. We compensate the reduced magnetic moment by increasing the magnetic-field gradient to \SI{40}{G/cm} in about \SI{200}{\mu s}. This levitates the Feshbach molecules against gravity and, in addition, removes unassociated atoms from the trap due to their larger magnetic moment. 

Imperfect timing of the magnetic-field and magnetic-field-gradient ramps results in a temporary tilt of the trap which induces collective oscillations of the molecular cloud. We minimize the collective oscillations by carefully programming the magnetic field and subsequently the magnetic force on the molecules over the course of approximately \SI{25}{ms} once the magnetic-field gradient has reached \SI{40}{G/cm} (see Fig.~\ref{fig:sequence}c). Once the collective oscillations have stopped, the 1550/1064-nm trap is compressed within \SI{30}{ms}. Since the trap is deep enough to hold the molecules without magnetic-field levitation, the magnetic-field gradient is ramped down within \SI{30}{ms} and the magnetic field is ramped to \SI{72.3}{G} where the magnetic moment of Feshbach molecules vanishes. Finally, the stimulated Raman adiabatic passage (STIRAP) scheme is employed, as described in Ref.~\cite{Park2015}.

\begin{figure}
\centering
\includegraphics{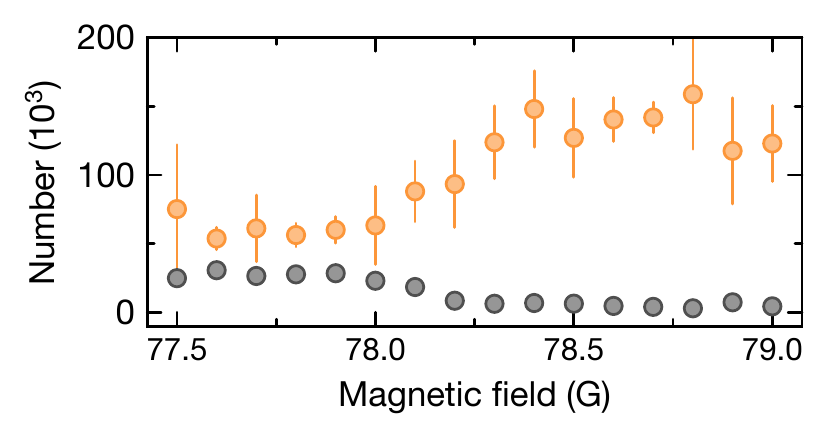}
\caption{Feshbach association in the 1550/1064-nm trap. Number of Feshbach molecules (gray) and the sum of associated and unassociated sodium atoms (orange).}
\label{fig:no-compression}
\end{figure}

\subsection{Feshbach association in 1550/1064-nm trap}
As a comparison to the species-dependent 785-nm trap, we characterize the association of Feshbach molecules in our typical 1550/1064-nm trap, where the densities are strongly mismatched as shown in Fig.~1c of the main text. The measurement starts with $2.9 \times 10^5$ $\mathrm{K}$ atoms and $1.4 \times 10^5$ $\mathrm{Na}$ atoms where the condensed fraction of the Na atoms is around $60 \%$. In contrast to the measurements described above, the ramp speed of the magnetic field across the Feshbach resonance is changed to \SI{2.4}{G/ms} to maximize the number of Feshbach molecules. The association ramp through the Feshbach resonance is stopped at different magnetic fields followed by a magnetic-field jump to \SI{72.3}{G} for Stern--Gerlach detection. Fig.~\ref{fig:no-compression} shows that only 30000 Feshbach molecules are associated resulting in an association efficiency of less than $25 \%$. Around $60 \%$ of the Na atoms, primarily Na atoms in the BEC, are lost due to interspecies collisions. This observation is in contrast to the measurements in the 785-nm trap where at most 10\% of the Na atoms are lost.

\subsection{Ramp speed of the Feshbach association}
\begin{figure}
\centering
\includegraphics{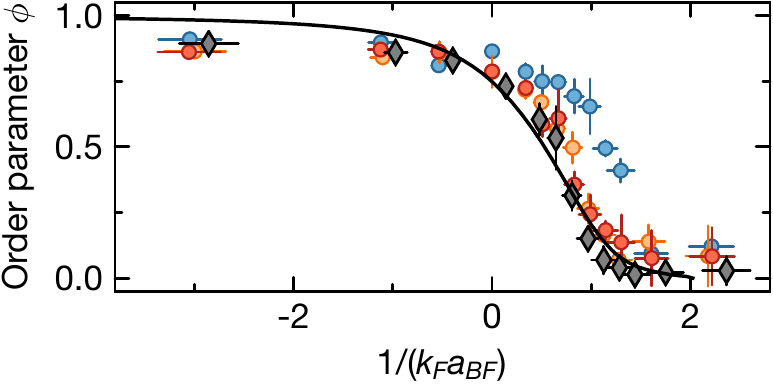}
\caption{Feshbach association at various ramp speeds. Order parameter $\phi$ as a function of $(k_F a_{BF})^{-1}$ at ramp speeds of \SI{1.7}{G/ms} (red points),  \SI{3.5}{G/ms} (orange points) and \SI{9}{G/ms} (blue points). In addition, we show the data from the main text at the ramp speed of \SI{3.5}{G/ms} for $\tilde{n}_B/\tilde{n}_F=1.3$ (grey diamonds).}
\label{fig:rampspeed}
\end{figure}

We measure the depletion of the condensate fraction as a function of $(k_F a_{BF})^{-1}$ at three different ramp speeds.  As shown in Fig.~\ref{fig:rampspeed}, the depletion of the condensate coincides for ramp speeds of \SI{1.7}{G/ms} and \SI{3.5}{G/ms} while showing a shift towards higher $(k_F a_{BF})^{-1}$ for ramp speeds of \SI{9}{G/ms}. We attribute the higher residual condensate fraction for the data shown by points to an imperfect trap alignment which leads to a worse density matching between bosons and fermions.

\subsection{Comparison of NSCT and fRG calculations}
The non-self-consistent T-matrix (NSCT) approach, from which we obtain the condensate depletion at equal density, predicts the polaron-to-molecule transition to occur at $(k_F a_{BF})^{-1}=1.60$ \cite{Guidini2015}. For finite boson densities it predicts  the phase transition between the polaronic condensate and the molecular phase to take place beyond this value as the boson concentration increases, i.e. $(k_F a_{BF})_c^{-1}>1.60$. Specifically, for the case of balanced densities, $n_B=n_F$, it predicts $(k_F a_{BF})_c^{-1}=2.02$. 

The NSCT approach, however, only takes into account single particle-hole excitations of the Fermi sea \cite{Lobo2006} and underestimates the modification of the binding energy of molecules inside the many-body environment. Indeed, when applying the NSCT approach  \cite{Guidini2015} to the Fermi-polaron problem at mass balance one finds the polaron-to-molecule transition to occur at $(k_F a_{BF})_c^{-1}=1.27$ \cite{Punk2009}, while more accurate techniques that include higher-order correlations such as functional renormalization group (fRG) \cite{Schmidt2011}  and state-of-the-art diagramatic Monte Carlo (QMC)  \cite{Prokofiev2009a,Prokofiev2009b}  predict a value of $(k_F a_{BF})_c^{-1}=0.90$.

Thus, in order to obtain a more accurate description of the critical interaction strength for the heteronuclear case considered in this work, we employ an fRG scheme which takes into account an infinite number of particle-hole excitations in the Fermi sea \cite{Schmidt2011}. The resulting polaron and molecule energies are shown in Fig.~1e of the main text, yielding a polaron-to-molecule transition at $(k_F a_{BF})^{-1}=1.16$. As shown in Fig.~1d, we find that the polaron quasiparticle weight obtained in the impurity limit already describes the condensate fraction well except for its discontinuity at the polaron-to-molecule transition. From investigations of two-component Fermi gases it is expected that this discontinuity will  be smoothed out due to finite boson density, temperature or combinations thereof \cite{Ness2020,Parish2021}. Moreover,  as explicitly shown in \cite{Guidini2015,Ness2020,Parish2021,vonMilczewski2021} and suggested by mean-field arguments \cite{vonMilczewski2021}, one expects the transition to shift to larger values of $(k_F a_{BF})^{-1}$ as the boson density increases. Hence, the value of $(k_F a_{BF})^{-1}=1.16$ obtained from the fRG in the impurity limit can be regarded as a lower bound on the actual location of the quantum phase transition at $(k_F a_{BF})_c$.

\begin{figure}
\centering
\includegraphics{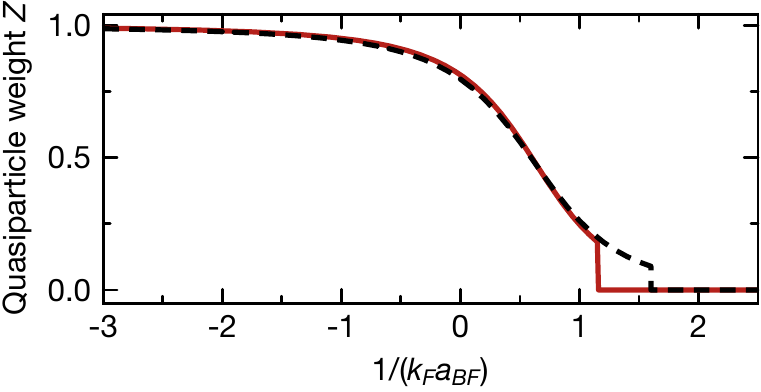}
\caption{Impurity quasiparticle weight in the Fermi polaron problem. The quasiparticle weight of the impurity as a function of the interaction strength is shown as obtained from the fRG (red, solid) and the NSCT approach (black, dashed). While both methods yield similar results  they differ in  the prediction of the point where the polaron-to-molecule transitions occurs, beyond which the occupied quasiparticle weight drops to zero.}
\label{fig:z_nsct_fRG}
\end{figure}

We note that the polaron and molecule energies cross at a rather shallow angle (see Fig.~1e in the main text). As a result, the underestimation of the molecule energy is the main reason for the difference in the predicted location of the polaron-to-molecule transition in the NSCT and fRG calculations. The quasiparticle weight is, in contrast, less affected and, as shown in Fig.~\ref{fig:z_nsct_fRG}, both approaches yield similar results for the quasiparticle weight of a bosonic impurity. Based on this finding, NSCT theory can be expected to give a reliable prediction for the condensate fraction also at finite boson density~\cite{Guidini2015}.

\subsection{Projection of polaronic states onto deeply bound molecules}
As evident from Fig.~2b in the main text, we detect a finite number of Feshbach molecules already before the phase transition. This can be understood from the fact that the rapid ramp to \SI{72.3}{G} projects the system onto deeply bound molecules. As a result, short-distance pairing correlations between bosons and fermions are effectively measured before the phase transition. However, once the phase transition is crossed in the initial magnetic-field ramp, all bosons are bound into weakly bound molecules in an adiabatic fashion. In that case, the weakly bound Feshbach molecules are transferred into deeply bound states by the subsequent rapid magnetic-field ramp.

\begin{figure}
\centering
\includegraphics{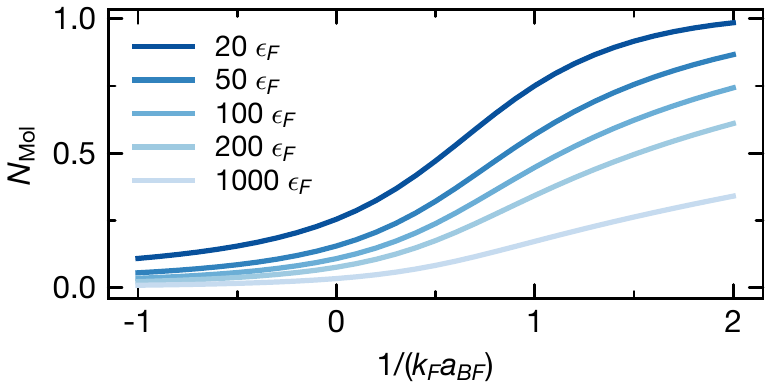}
\caption{Number of projected molecules. The projected molecule fraction $N_{\text{Mol}}$ obtained from  the Fermi polaron state within the Chevy Ansatz \cite{Chevy2006} is shown as a function of the interaction parameter $(k_F a_{BF})^{-1}$ for different binding energies of the deeply bound molecule.}
\label{fig:PolProjection}
\end{figure}

Here we demonstrate that a simple Fermi-polaron wave function indeed overlaps with deeply bound molecules when a projection measurement as described in the main text is performed. Similar to the calculations performed in Ref.~\cite{Pini2020} we define the molecule number operator as 
\begin{align}
    \hat{N}_{\text{Mol}}= \sum_{\mathbf{mlk}} c^\dagger_{\mathbf{m}+\mathbf{k}} d^{\dagger}_{-\mathbf{k}} \phi^{\phantom{*}}_\mathbf{m}(\mathbf{k}) \phi^{*}_\mathbf{m} (\mathbf{l} ) d^{\phantom{\dagger}}_{-\mathbf{l}}c^{\phantom{\dagger}}_{\mathbf{m}+\mathbf{k}} \ . 
\end{align}
Here, $d_\mathbf{q}$ and $c_\mathbf{q}$ are the fermionic and bosonic annihilation operators, respectively, while $\phi_{\mathbf{p}} (\mathbf{k})$ denotes the  wave function of the molecule in vacuum at center-of-mass momentum $\mathbf{p}$. We approximate it with the form valid for an attractive contact interaction potential, 
\begin{align}\label{eqmolfunc}
    \tilde{\phi}_{\mathbf{p}}(\mathbf{k})&= \frac{1}{E_b + \frac{\alpha+1}{2 m \alpha} \left(\mathbf{k}+ \frac{\alpha}{1+\alpha} \mathbf{p}\right)^2},\\
   \phi_{\mathbf{p}}(\mathbf{k})&=  \frac{\tilde{\phi}_{\mathbf{p}}(\mathbf{k})}{\sqrt{\sum_{\mathbf{l}}  \left|\tilde{\phi}_{\mathbf{p}}(\mathbf{l})\right|^2  }}.
\end{align}
 Here $E_b$ denotes the energy of the  molecule at $\mathbf{p}=\mathbf{0}$, $m$ denotes the mass of the fermions and $\alpha$ is given by the ratio of the bosonic and the fermionic mass. 
 
The operator $\hat N_\text{Mol}$  measures the number of projected molecules with respect to the Chevy Ansatz \cite{Chevy2006} for the Fermi polaron
\begin{align}
    \ket{\text{Pol}}= \alpha^{\phantom{\dagger}}_0 d^{\dagger}_{\mathbf{0}}\ket{\text{FS(N)}}+ \sum_{\mathbf{kq}} \alpha^{\phantom{\dagger}}_{\mathbf{kq}} d^\dagger_{\mathbf{q}-\mathbf{k}} c^\dagger_{\mathbf{k}}c^{\phantom{\dagger}}_{\mathbf{q}} \ket{\text{FS(N)}}
\end{align}
as $N_{\text{Mol}}=\bra{\text{Pol}} \hat{N}_{\text{Mol}} \ket{\text{Pol}}$. Here $\alpha_0$ and $\alpha_{\mathbf{kq}}$ denote variational parameters and $\ket{\text{FS(N)}}$ denotes a Fermi sea containing $\text{N}$ fermions. 

The resulting fractions are shown in Fig.~\ref{fig:PolProjection} for a mass ratio of $\alpha=23/40$ and different values of $E_b$. For the binding energies shown, it can be seen that with increasing $(k_F a_{BF})^{-1}$ the molecule fraction increases, with the fraction getting smaller as  $E_b$ increases. This shows that, although the polaron is by no means a molecular state, it still features pairing correlations that will lead to a finite overlap with deeply bound molecules. Note that when the transition to the molecular phase is reached, in our experimental procedure the associated weakly bound molecules will be transferred nearly adiabatically to the more deeply bound molecules approximately described by Eq.~\eqref{eqmolfunc}.  Thus the number of observed molecules would approximately saturate (in absence of losses). Since during the final part of the initial ramp, however, excess bosons  undergo lossy collisions with the molecules, the number of molecules will be further reduced and it is thus expected that the number of observed molecules will in fact be maximized at the transition using our experimental procedure.

\subsection{Two-body loss of Feshbach and ground-state molecules}
We measure the two-body loss rate of Feshbach and ground-state molecules. For Feshbach molecules, the loss measurement is performed after the 100-ms holding time during which collective oscillations are dampened out and the 1550/1064-nm trap has been compressed. The loss measurement of ground-state molecules is performed immediately after the transfer into the ground state followed by a removal of the residual Feshbach molecules by resonant light pulses. The trapping frequency for Feshbach and ground-state molecules is $(\omega_x,\omega_y,\omega_z)= 2 \pi \times (71,99,234)$ Hz because we increase the power of the 1550/1064-nm trap to compensate for the change of polarizabilities between the Feshbach and the ground-state molecules. 

\begin{figure}
\centering
\includegraphics{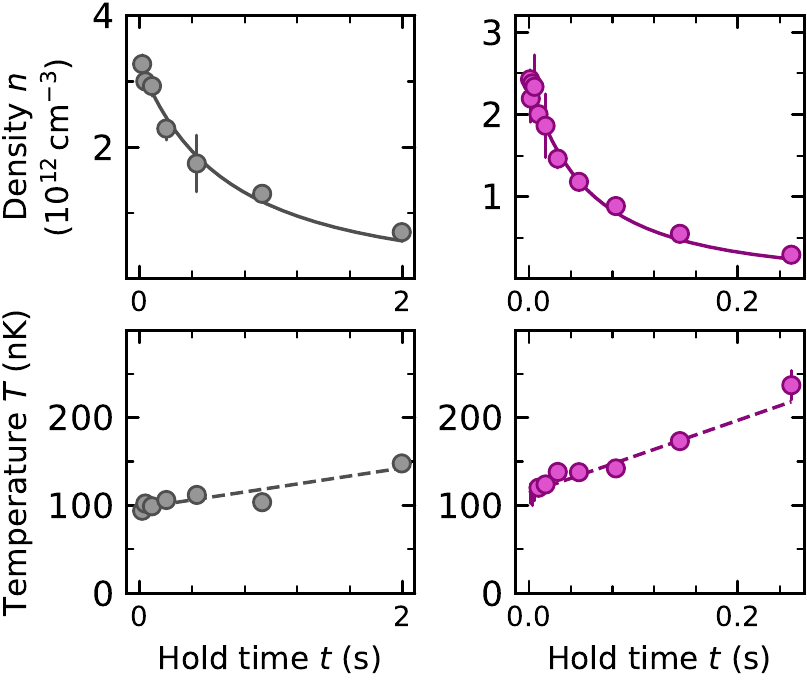}
\caption{Lifetime of $\mathrm{NaK^*}$ Feshbach (left panel) and $\mathrm{NaK}$ ground-state molecules (right panel). The upper panel figure shows the average density $n$ of the Feshbach molecules (gray points) and ground-state molecules (purple points) as a function of holding time. The data is fitted with Eq.~\ref{eq:density-loss-closedform}. The result of the fits are plotted as solid lines. The error bars are statistical errors derived from the atom number fluctuations. For the fitting procedure we only consider data points where the molecule number is larger than $1 \times 10^4$. The lower panel shows the temperature of the Feshbach molecules (gray points) and the ground-state molecules (purple points) as a function of the holding time extracted from Fermi-Dirac distribution fits. The dashed lines represent the linear fit of the temperature.}
\label{fig:loss}
\end{figure}

We start from a density-dependent two-body loss model with heating
\begin{align}
\label{eq:density-loss}
\frac{\mathrm{d}n}{\mathrm{dt}} = -\beta n^2 - \frac{3}{2} \frac{n}{T} \frac{dT}{dt}   \, ,
\end{align}

where $n$ represents the average density and $T$ the temperature of the molecular sample. The two-body loss coefficient is expected to scale linearly with temperature which was experimentally confirmed for fermionic ground-state molecules \cite{Ospelkaus2010,Bause2021}. With this knowledge we define the temperature-independent loss coefficient $b = \beta / T$, where we obtain the temperature $T$ by fitting Fermi-Dirac distribution to the molecular clouds. In a first step, we fit the temperature with a linear-heating model $T_{\mathrm{lin}}(t)=T_0+h t$, where $T_0$ is the initial temperature and $h$ the heating rate. Using this linear model for heating, we solve Eq.~\eqref{eq:density-loss} and obtain \cite{DeMarco2019} 

\begin{widetext}
\begin{equation}
\label{eq:density-loss-closedform}
n(t) = \frac{n_0 h T_0^{3/2}}{T_{\mathrm{lin}}(t)\left(2 n_0 T_0^2(\sqrt{T_0}-\sqrt{T_{\mathrm{lin}}(t)})b + h (\sqrt{T_{\mathrm{lin}}(t)}+2 n_0 t T_0^{3/2}b)\right)} \, ,
\end{equation}
\end{widetext}

where $n_0$ describes the initial density of the sample. We fit Eq.~\eqref{eq:density-loss-closedform} to our data with $n_0$ and $b$ as free parameters. In analogy to Ref.~\cite{DeMarco2019}, we extract the density of the sample by approximating the degenerate Fermi gas as a classical gas at a temperature of $T_{\mathrm{cl}}$. We obtain $T_{\mathrm{cl}}$ by fitting a Gaussian distribution to the time-of-flight images. The average in-situ density of the classical gas is given by
\begin{align}
\label{eq:density}
n(T_{\mathrm{cl}}) = \frac{N}{8 \pi^{3/2}}\bar{\omega}^3 \left( \frac{k_B T_{\mathrm{cl}}}{m} \right) ^{-3/2}, 
\end{align}

where $\bar{\omega}$ is the geometric average of the trapping frequency and the classical temperature $T_{\mathrm{cl}}$ accounts for the Fermi pressure as well as the kinetic energy of the sample, thus overestimating the actual temperature. Still, this effective model allows us to determine the density of the sample even in the degenerate regime with the least assumptions with respect to the loss and whether the sample is in thermal equilibrium. The accuracy in determining the density for degenerate Fermi samples using a Gaussian distribution instead of a Fermi--Dirac distribution was checked numerically in Ref.~\cite{DeMarco2001}. It shows that the error is negligible for Fermi gases at temperatures as low as $0.2~T_F$. Fig.~\ref{fig:loss} shows the loss measurement of the Feshbach and ground-state molecules. For Feshbach molecules we get $b_{\mathrm{FB}}=\SI{3.8(5)e-12}{cm^3 / \mu Ks}$ and for ground-state molecules we get $b_{\mathrm{GS}}= \SI{4.8(6)e-11}{cm^3 / \mu Ks}$.

\subsection{Error bars for the measurements}
The error bars in $(k_F a_{BF})^{-1}$ in Fig.~1d and Fig.~2 result from the magnetic field uncertainty of 15 mG and an error of 10 \% in the average trapping frequency. The vertical error bars in Fig.~1d, Fig.~2, Fig.~5, Fig.~6, and Fig.~9 are the standard deviation of 3-5 measurements.

\end{document}